\documentclass[aps, prd, amsmath, floats, floatfix, twocolumn, nofootinbib, showpacs]{revtex4-1}
\usepackage{graphicx}
\usepackage{color}
\usepackage{latexsym}

\newcommand{\be}{\begin{eqnarray}}
\newcommand{\ee}{\end{eqnarray}}
\newcommand{\rar}{\rightarrow}

\begin{document}

\title{X-ray spectropolarimetric measurements of the Kerr metric}

\author{Dan Liu}

\author{Zilong Li}

\author{Yifan Cheng}

\author{Cosimo Bambi}
\email[Corresponding author: ]{bambi@fudan.edu.cn}

\affiliation{Center for Field Theory and Particle Physics and Department of Physics, Fudan University, 200433 Shanghai, China}

\date{\today}

\begin{abstract}
It is thought that the spacetime geometry around black hole candidates is described by the Kerr solution, but an observational confirmation is still missing. Today, the continuum-fitting method and the analysis of the iron K$\alpha$ line cannot unambiguously test the Kerr paradigm because of the degeneracy among the parameters of the system, in the sense that it is impossible with current X-ray data to distinguish a Kerr black hole from a non-Kerr object with different values of the model parameters. In this paper, we study the possibility of testing the Kerr nature of black hole candidates with X-ray spectropolarimetric measurements. Within our simplified model that does not include the effect of returning radiation, we find that it is impossible to test the Kerr metric and the problem is still the strong correlation between the spin and possible deviations from the Kerr geometry. Moreover, the correlation is very similar to that of other two techniques, which makes the combination of different measurements not very helpful. Nevertheless, our results cannot be taken as conclusive and, in order to arrive at a final answer, the effect of returning radiation should be properly taken into account.
\end{abstract}


\maketitle


\section{Introduction \label{s-1}}

General relativity is today the best framework for the description of the gravitational force and the geometrical structure of the spacetime around massive bodies. In the past 60 years, the theory has passed a number of experimental tests and its predictions have been verified in the weak gravitational fields, mainly with precise experiments in the Solar System and accurate radio observations of binary pulsars~\cite{will}. Now the interest is to check the validity of the theory in more extreme environments. One of the most fascinating predictions of general relativity is the existence of black holes (BHs). In 4-dimensional general relativity, an uncharged BH is described by the Kerr solution and it is completely characterized by only two parameters, namely its mass $M$ and its spin parameter $a_* = a/M = J/M^2$, where $J$ is the BH spin angular momentum. Kerr BHs are thought to be the final stage of any heavy star after it exhausts all its nuclear fuel~\cite{collasso}. Initial deviations from the Kerr solution can indeed be quickly radiated away through the emission of gravitational waves~\cite{kerr1}. Any non-vanishing electric charge would be soon neutralized because of the highly ionized host environment of these objects~\cite{kerr2}. Deviations from the Kerr metric produced by the presence of accretion disks are normally completely negligible, as the disk mass is typically many orders of magnitude smaller than the mass of the central body~\cite{kerr3}.

Astrophysical BH candidates are dark and compact objects that can only be interpreted as Kerr BHs in the framework of conventional physics and they can be something else only in the presence of new physics. For instance, a compact object in an X-ray binary is classified as a BH candidate if its mass exceeds 3~$M_\odot$, because the latter is the maximum mass for a neutron star for any reasonable matter equation of state~\cite{rr}. Present observations cannot confirm that the spacetime geometry around BH candidates is really described by the Kerr metric~\cite{review}. For the time being, there are only two relatively robust techniques to study the nature of BH candidates, namely the analysis of the thermal spectrum of thin accretion disks (continuum-fitting method)~\cite{cfm} and the iron K$\alpha$ line~\cite{iron}. Both techniques have been developed to estimate the spin parameter under the assumption of the Kerr background and, more recently, they have been extended to test the nature of BH candidates~\cite{cfm2,code,iron2}.

The typical problem to verify the Kerr BH hypothesis is the degeneracy among the parameters of these systems, and in particular among the estimate of the spin, possible deviations from the Kerr geometry, and the inclination angle of the disk. The disk's thermal spectrum has a simple shape and therefore it is fundamentally impossible to distinguish the effect of the spin from non-Kerr metric elements~\cite{lingyao}. The iron line has a more complicated structure and, in the presence of the correct astrophysical model, it would be possible to distinguish Kerr and non-Kerr BHs with high quality data. However, this is impossible with current X-ray facilities, even in the case of very good observations~\cite{jjc}. Moreover, some kinds of deformations are definitively more difficult to constrain than others~\cite{jjc2}. With current data of the iron line, we can only rule out some BH alternatives without horizon, like some boson stars and some traversable wormholes~\cite{exotic}, because their iron line profile would have qualitatively different features. We note that the combination of the measurement of the continuum and of the iron line of the same object can unlikely break the degeneracy, because both techniques are mainly sensitive to the position of the inner edge of the disk~\cite{cfmiron}, which is normally supposed to be at the radius of the innermost stable circular orbit (ISCO) and it is thus only determined by the background metric\footnote{We note that it is often assumed that the emission of the plunging gas inside the ISCO can be completely neglected.}. The studies of quasi-periodic oscillations and of jet power are potentially other techniques to probe the metric around BH candidates~\cite{qpojet}, but the exact mechanism responsible for these phenomena is not yet well understood and therefore they cannot yet be used to test fundamental physics.

The study of the polarization of the thermal radiation of thin accretion disks may be a new technique to test the Kerr metric of stellar-mass BH candidates in the near future\footnote{Here we are interested in stellar-mass BH candidates because the disk temperature scale as $M^{-1/4}$ and in the case of a compact object of 10~$M_\odot$ the peak of the spectrum is around 1~keV. For supermassive BH candidates of millions or billions Solar masses, the thermal spectrum of a thin disk falls in the UV/optical band, where dust absorption makes an accurate measurement impossible.}. Such a radiation is initially unpolarized, but it gets polarized at the level of a few percent due to Thomson scattering of X-ray radiation off free electrons in the disk's atmosphere. In the Kerr metric, the degree and the angle of polarization depend on the BH spin and the inclination angle of the disk with respect to the line of sight of the observer~\cite{stark}. Assuming the Kerr background, spectropolarimetric observations could provide and estimate of these two parameters~\cite{lixin,schnittman}. While some polarimetric missions have been cancelled, similar measurements will be hopefully possible in the near future with the Chinese X-ray Timing Polarimetric (XTP) satellite~\cite{xtp}, which may be launched in 2020, and the European X-ray Imaging Polarimetry Explorer (XIPE) mission~\cite{xipe}, which was recently approved for initial design work.

X-ray spectropolarimetric measurements can be potentially used to test the Kerr metric. Such a possibility has been already explored in Ref.~\cite{kraw}, where is was found that these measurements are mainly sensitive to the position of the ISCO. In the present paper, we want to perform a more detailed analysis on the correlation among the parameters of the system and on future detection capabilities. As an explorative study in this direction, we make some simplifications, and in particular we neglect the returning radiation, namely the radiation emitted by the disk that comes back to the disk because of the strong light bending in the vicinity of the compact object. First, we confirm the strong correlation between the spin and possible deviations from the Kerr geometry. Second, the correlation is very similar to that found in the estimates with the continuum-fitting method and the iron line analysis, which suggests that the possible combination of the three measurements is not promising to break the degeneracy. Third, considering some plausible future measurements of the polarization of the continuum, we find that constraints with this method are not better than those from the thermal spectrum of thin disks and surely worse than those from the iron line. Even in the case of objects that look like fast-rotating Kerr BH observed from quite a large inclination angle, it is impossible to exclude significant deviations from the Kerr metric. Fourth, if we do not assume {\it a priori} the Kerr background, we lose also the capability of obtaining stringent constraints on the inclination angle of the disk.

We note that our results are not conclusive, because we have neglected the effect of returning radiation. The latter depends on the light bending in the vicinity of the BH and it is possible that its signature improves the measurement of the Kerr metric, even because it is not strictly related to the position of the ISCO radius. As clearly shown in Ref.~\cite{schnittman}, the signature imprinted by the returning radiation is weak below 10~keV for a Schwarzschild BH (roughly speaking, when the ISCO radius is not too close to the compact object). However, it becomes prominent for a fast-rotating BH from energies above a few keV, which is an energy range covered by the proposed polarimetric missions that should do measurements at $\sim$1-10~keV. The effect will be included in a future work, since it cannot be treated as in the Kerr background and the calculations become very time consuming.

The content of this paper is as follows. In the next section, we briefly review our theoretical model for the description of the accretion disk and the spacetime metric of a BH candidate. In Section~\ref{s-3}, we describe our calculations of the spectrum of the polarization degree and the polarization angle. In Section~\ref{s-4}, we study the degeneracy among the parameters of our model, namely the spin parameter, the deformation parameter, and the inclination angle of the disk with respect to the line of sight of the distant observer. Summary and conclusions are reported in Section~\ref{s-c}. In the appendix, we provide some details about the calculations of the polarization of the radiation from the disk, since we cannot use the usual approach adopted in the Kerr metric exploiting the nice properties of the Kerr geometry. Our approach is more general, in the sense it can be applied to any stationary, axisymmetric, and asymptotically flat spacetime, but it is inevitably more time consuming. Throughout the paper, we use units in which $G_{\rm N} = c = 1$.

\section{Theoretical model} \label{s-2}

\subsection{Accretion disk}

We describe the accretion disk with the Novikov-Thorne model~\cite{nt-model}, which is the standard set-up for geometrically thin and optically thick accretion disks in stationary and axisymmetric spacetime~\cite{cfm}. The disk is assumed to be on the equatorial plane, the particles of the gas follow nearly geodesic circular orbits, and the inner edge of the disk is at the ISCO radius. See e.g. the last paper in~\cite{cfm} for the validity of these assumptions. From the conservations laws for rest-mass, energy, and angular momentum, we can derive the time-averaged structure of the disk. The disk is in thermal equilibrium and the emission at any radius is like that of a blackbody, so that we can define an effective temperature as a function of the radial coordinate, $T_{\rm eff}(r)$. Since the temperature in the inner part of the accretion disk around a stellar-mass BH candidate can be very high, up to $\sim 10^7$~K, electron scattering in the atmosphere makes the spectrum deviate from the original Novikov-Thorne prediction. This can be taken into account by introducing the color factor (or hardening factor) $f_{\rm c}$~\cite{shimura}. 
Since we are interested in stellar-mass BH candidates with an accretion luminosity of about 10\% the Eddington luminosity of the object, throughout the paper we use $f_c = 1.6$. The color temperature is defined as $T_{\rm c} = f_{\rm c} T_{\rm eff}$ and the local specific intensity of the radiation emitted by the disk is
\be
I(\nu_{\rm e}) = \frac{1}{f_{\rm c}^4} B (T_{\rm c}) \Upsilon \, ,
\ee
where $\nu_{\rm e}$ is the photon frequency in the rest frame of the gas, $B(T_{\rm c})$ is the blackbody function for the temperature $T = T_{\rm c}$, and $\Upsilon$ is a function of the angle between the direction of the propagation of the photon and the normal to the disk surface.

\begin{figure}
\begin{center}
\includegraphics[type=pdf,ext=.pdf,read=.pdf,width=8.7cm]{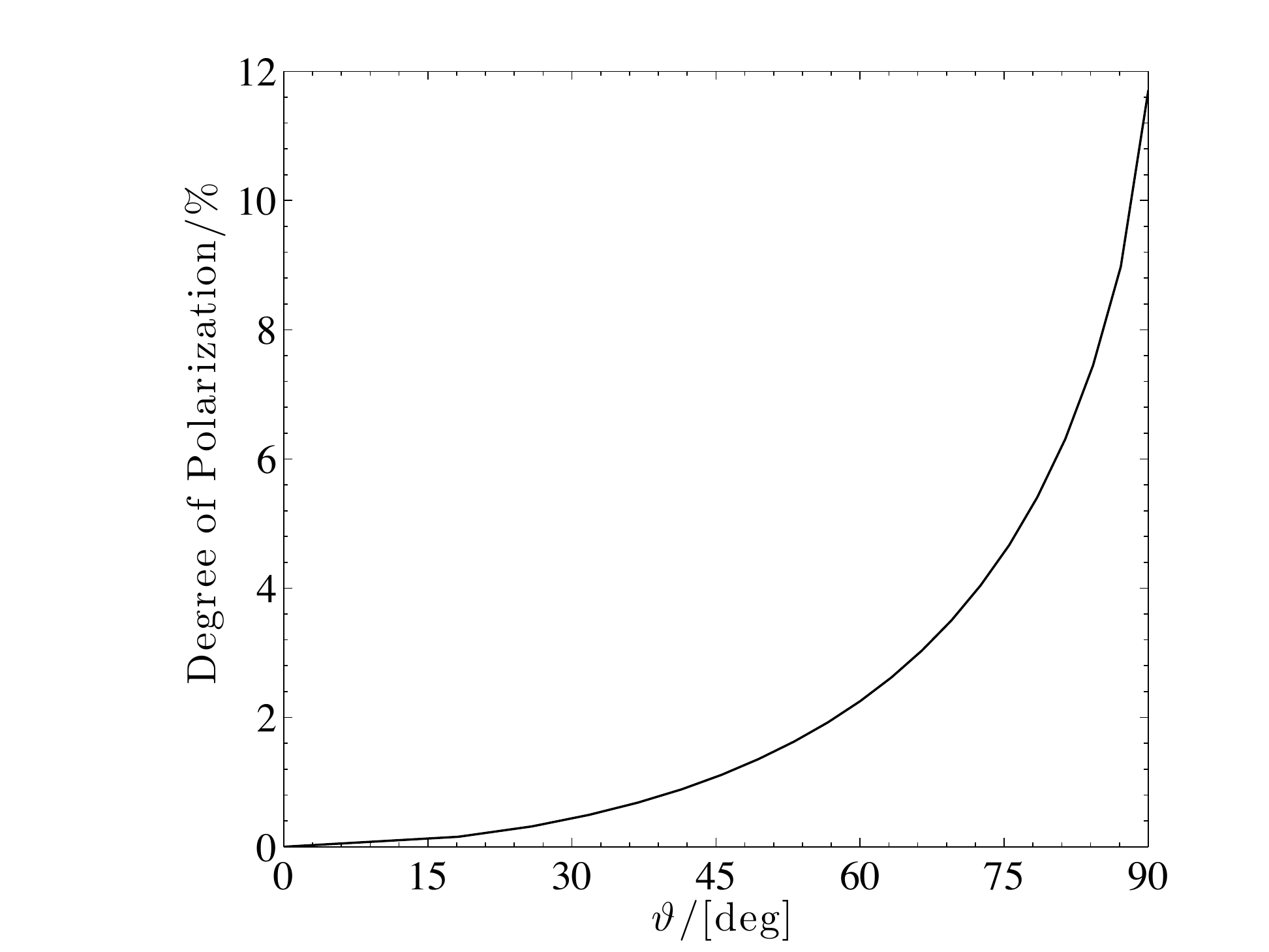}
\end{center}
\caption{Polarization degree as a function of the angle $\vartheta$ between the normal to the disk surface and the direction of propagation of the photon. }
\label{fig0}
\end{figure}

Thermal radiation is initially unpolarized. However, because of the Thomson scattering of photons off free electrons in the dense atmosphere of the disk, the radiation becomes partially polarized. With reference to the rest frame of the gas, the degree of polarization depends on the angle $\vartheta$ between the normal to the disk surface and the direction of propagation of the X-ray photon, ranging from 0 ($\vartheta = 0^\circ$, photon direction parallel to the normal to the disk) to about 12\% ($\vartheta = 90^\circ$, photon direction orthogonal to the normal to the disk)~\cite{chandra}. The polarization degree as a function of $\vartheta$ (which is measured in the rest frame of the gas) is shown in Fig.~\ref{fig0}. The orientation of the polarization vector is instead parallel to the disk plane and orthogonal to the direction of propagation of the photon. The same scattering in the disk atmosphere causes a limb-darkened emission and the correct value for $\Upsilon$ can be found in the table in~\cite{chandra}. More details can be found in~\cite{lixin,schnittman}.

In the present paper, we adopt two main simplifications in the calculation of the polarization. We do not take into account the effect of the returning radiation, namely the effect of the photons that are emitted by the disk and, because of the strong gravitational field near the compact object, return to the disk, increasing its temperature and local specific intensity~\cite{schnittman}. Such an effect is clearly more important in the high energy part of the spectrum, because high energy photons are produced at small radii, and for fast-rotating BHs, because the inner edge of the disk is closer to the compact object. In the latter case, the effect can indeed be important just above a few keV (for more details, see Ref.~\cite{schnittman}) and therefore our results cannot be taken as conclusive.

The second ingredient that is neglected in our calculation is the photon absorption, which tends to destroy the polarization and it is more important for low energy photons, so at larger radii~\cite{lixin}. This effect is expected to be important below $\sim 0.5$~keV and therefore should not be relevant for the first generation of polarimetric detectors.

\subsection{Background metric}

In order to test the Kerr metric around BH candidates, it is necessary to constrain possible deviations from the Kerr geometry. Indeed, it is not enough that observational data nicely fit a Kerr model, because a non-Kerr object may look like a Kerr BH with different values of the model parameters. As example, we can consider the approach used in Solar System experiments. In this case, we want to check the Schwarzschild solution in the weak field limit. In the Parametrized Post-Newtonian (PPN) formalism~\cite{will}, we write the most general static and spherically symmetric metric as an expansion in $M/r$ 
\be\label{eq-ppn}
ds^2 &=& - \left(1 - \frac{2 M}{r} + \beta \frac{2 M^2}{r^2} + . . . \right) dt^2 \nonumber\\
&& + \left(1 + \gamma \frac{2 M}{r} + . . . \right) \left(dx^2 + dy^2 + dz^2\right) \, ,
\ee
where $\beta$ and $\gamma$ are two coefficients in the expansion and they parametrize our ignorance. In general relativity, the only spherically symmetric vacuum solution is the Schwarzschild metric and, when cast in the form above, we have $\beta = \gamma = 1$. To test the Schwarzschild metric, we employ the line element in Eq.~(\ref{eq-ppn}) and we determine $\beta$ and $\gamma$ from observations. Current data require that $\beta$ and $\gamma$ are 1 at the level of $10^{-5} - 10^{-4}$ and this confirms the Schwarzschild solution with this precision~\cite{will}.

The same strategy can be employed to test the Kerr metric. At present, there is not a satisfactory approach like the PPN formalism: since we want to probe the spacetime close to the compact object, we cannot use an expansion in $M/r$ and it is thus difficult to take into account any kind of deviations from the Kerr solution. In this paper, we adopt the Johansenn-Psaltis metric~\cite{jp-m}, which is a quite popular metric to test the Kerr spacetime. In Boyer-Lindquist coordinates, the line element reads~\cite{jp-m}
\begin{widetext}
\be\label{eq-jp}
ds^2 &=& - \left(1 - \frac{2 M r}{\Sigma}\right) \left(1 + h\right) dt^2
- \frac{4 a M r \sin^2\theta}{\Sigma} \left(1 + h\right) dtd\phi
+ \frac{\Sigma \left(1 + h\right)}{\Delta + a^2 h \sin^2 \theta} dr^2
+ \nonumber\\ &&
+ \Sigma d\theta^2 + \left[ \left(r^2 + a^2 +
\frac{2 a^2 M r \sin^2\theta}{\Sigma}\right) \sin^2\theta +
\frac{a^2 (\Sigma + 2 M r) \sin^4\theta}{\Sigma} h \right] d\phi^2 \, ,
\ee
\end{widetext}
where $\Sigma = r^2 + a^2 \cos^2\theta$ and $\Delta = r^2 - 2 M r + a^2$. In its simplest version, $h$ is given by
\be
h = \frac{\epsilon_3 M^3 r}{\Sigma^2} \, .
\ee
$\epsilon_3$ is the ``deformation parameter'' and it is used to quantify possible deviations from the Kerr geometry. The compact object is more prolate (oblate) than a Kerr BH for $\epsilon_3 > 0$ ($\epsilon_3 < 0$); when $\epsilon_3 = 0$, we exactly recover the Kerr solution.

\begin{figure*}
\begin{center}
\includegraphics[type=pdf,ext=.pdf,read=.pdf,width=8.7cm]{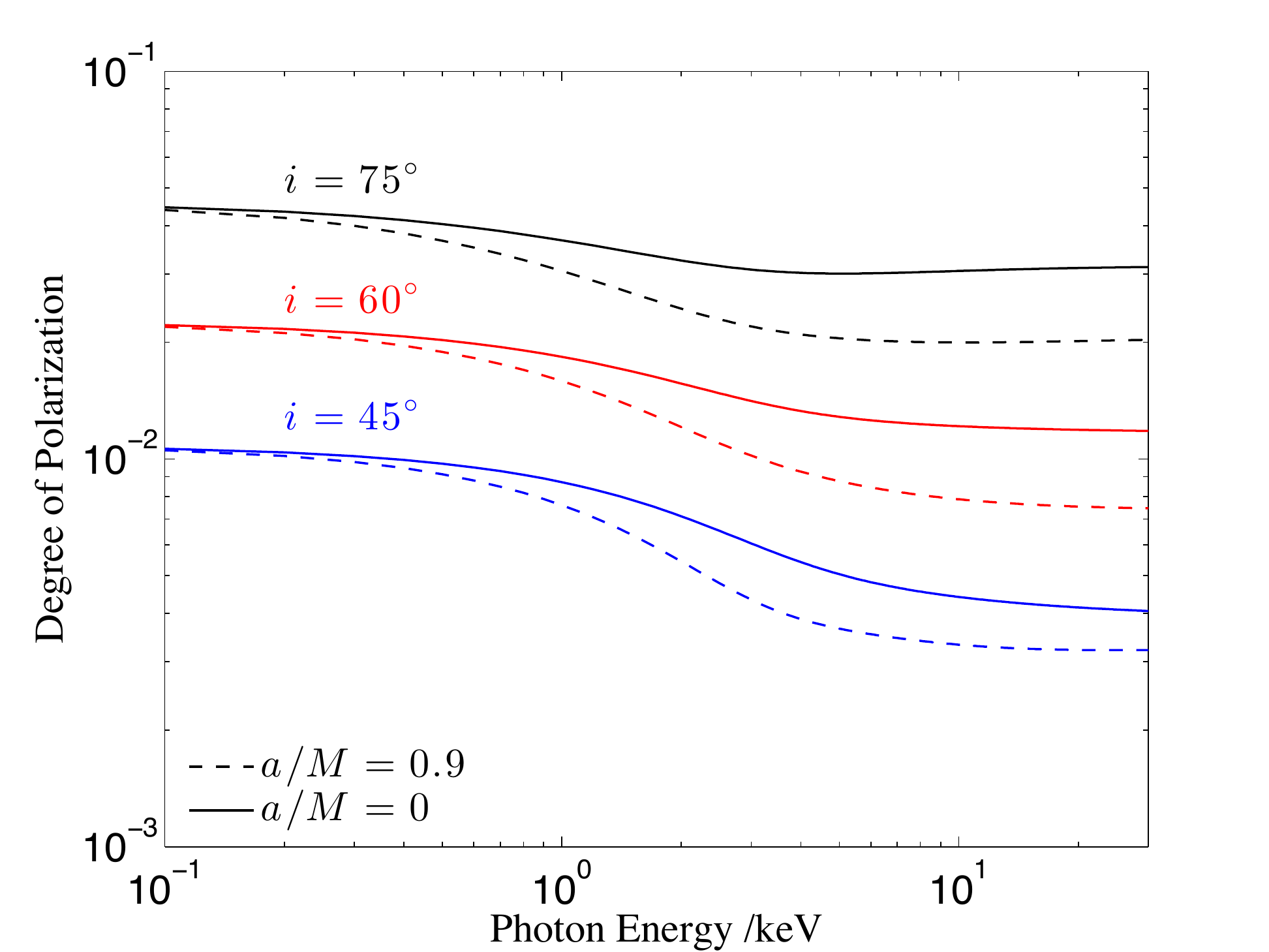}
\includegraphics[type=pdf,ext=.pdf,read=.pdf,width=8.7cm]{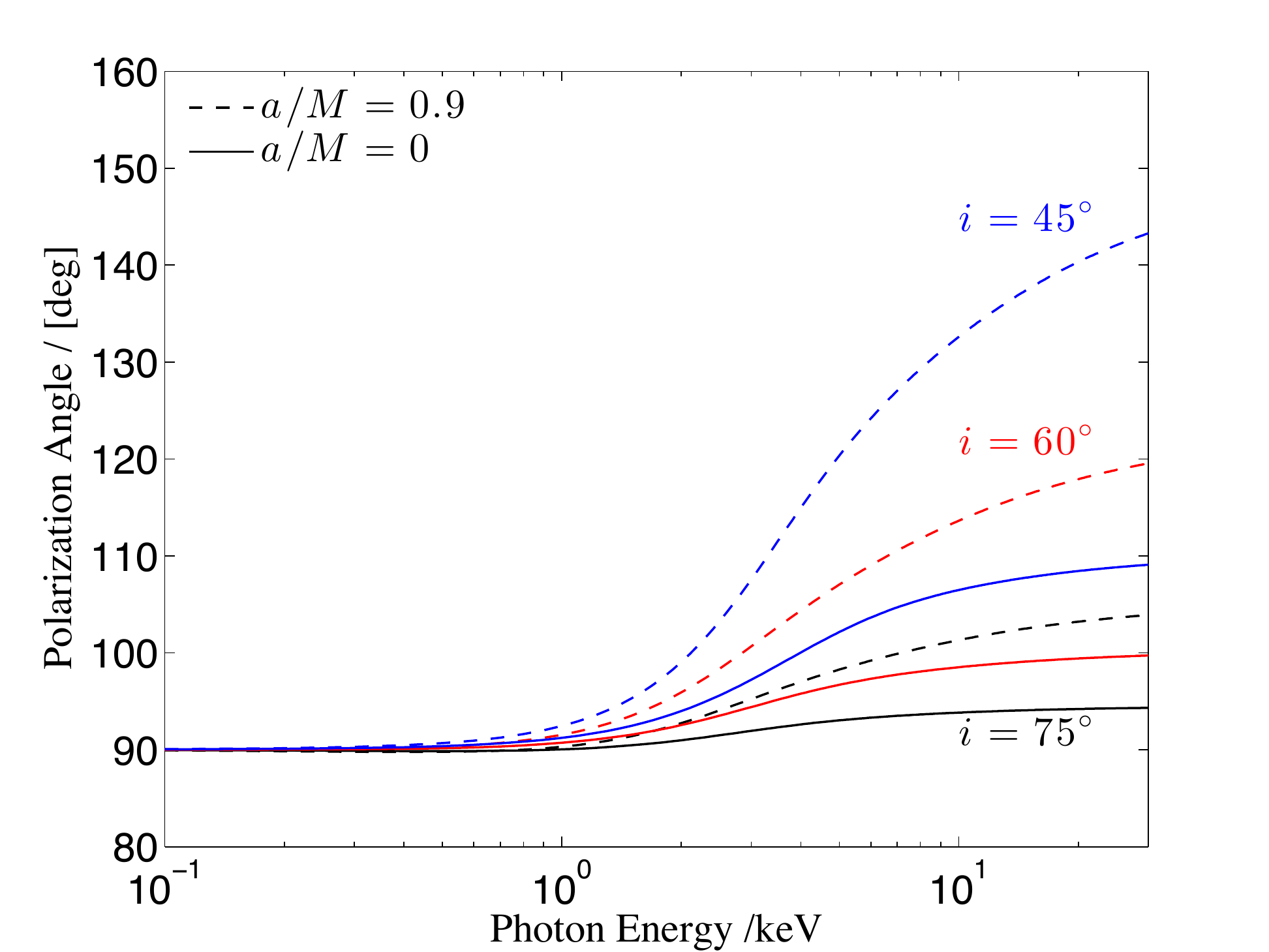}
\end{center}
\caption{Polarization degree (left panel) and polarization angle (right panel) as a function of the photon energy for a Schwarzschild BH (solid lines) and a Kerr BH with spin parameter $a/M = 0.9$ (dashed lines) and a viewing angle $i=45^\circ$, $60^\circ$, and $75^\circ$. See the text for more details.}
\label{fig1}
\end{figure*}

\begin{figure}
\begin{center}
\includegraphics[type=pdf,ext=.pdf,read=.pdf,width=9cm]{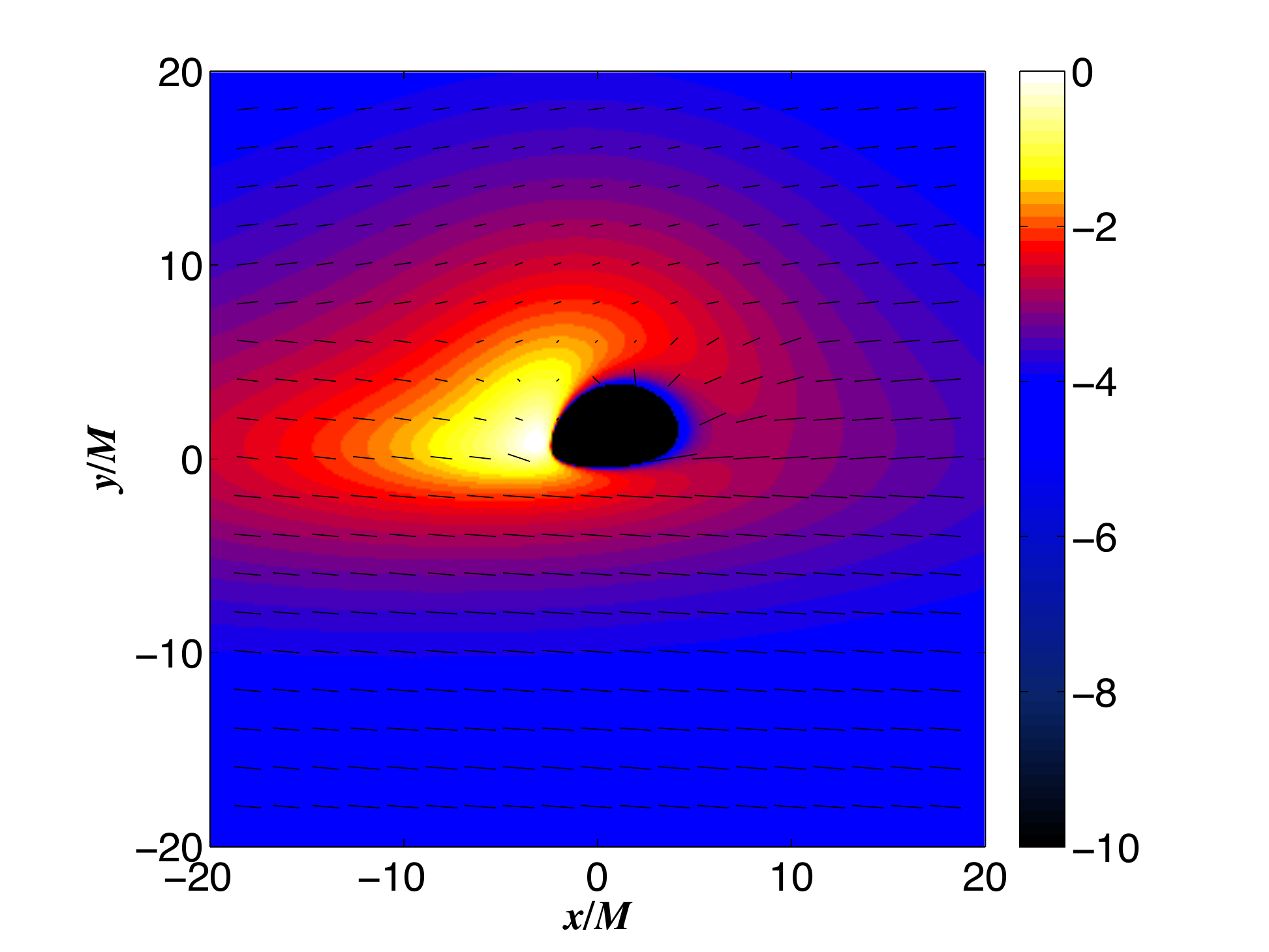}
\end{center}
\caption{Apparent image of the accretion disk around a Kerr BH with spin parameter $a/M = 0.99$ and observed from a viewing angle $i=75^\circ$. The contour map shows the total intensity of the radiation (logarithmic scale). The segments report the properties of the polarization of the radiation: the length of every segment is proportional to the polarization degree, while the orientation of every segment corresponds to the orientation of the polarization vector. See the text for more details.}
\label{fig2}
\end{figure}

\section{Numerical simulations} \label{s-3}

We use the code described in Ref.~\cite{code}, which has been extended to compute the degree and the angle of polarization of the thermal spectrum of a thin disk. We remind the reader that we are performing these calculations in a non-Kerr background, and therefore we cannot adopt the usual approach used in a Kerr code, in which one exploits the fact that the Kerr solution is a Petrov type~D spacetime. As discussed in~\cite{code}, the calculations of the photon propagation backward in time from the observer's plane to the point of the photon emission in the accretion disk is done by solving the geodesic equations. Now we also need to compute the polarization degree $\delta$ and the polarization angle $\psi$ for any photon on the observer's plane. The polarization degree is a scalar and it only depends on the angle $\vartheta$ between the normal to the disk surface and the direction of propagation of the X-ray photon and it does not require special prescription. For the polarization angle, we need to parallel transport the polarization vector along the photon geodesic; we cannot exploit the Walker-Penrose theorem valid for Petrov type~D spacetimes, and therefore we need to solve the basic equations for parallel transport. At any point of the grid of the observer's plane, we simultaneously compute the photon trajectory (backward in time) and the propagation of an auxiliary vector. When the photon reaches the disk, we evaluate the polarization degree and the angle difference between the propagated auxiliary vector and the polarization vector at the emission point. Since the angle between the two vectors is conserved along the geodesic, we can immediately determine the angle of the polarization vector on the plane of the distant observer. The details are given in the appendix at the end of this paper.

Once we know the polarization degree and angle at each point in the grid, we need to integrate over the observer's plane to get the spectrum of $\delta$ and $\psi$ (in this paper we use the notation of Ref.~\cite{lixin}). In terms of the Stokes parameters $I$, $Q$, $U$, and $V$~\cite{chandra}, for each point on the image we have
\be
Q + i U = \delta I e^{2 i \psi} \, ,
\ee
where $V=0$ because the radiation is linearly polarized. The radiation field is decomposed into a completely polarized component $I^p = \delta I$ and an unpolarized one $I^u = (1-\delta) I$. At the point of the detection, we have
\be
\vspace{-0.8cm}
\langle Q_{\rm obs} \rangle + i \langle U_{\rm obs} \rangle &=&
\frac{1}{\Delta \Omega_{\rm obs}} \int \left( Q_{\rm obs} 
+ i U_{\rm obs} \right) d\Omega_{\rm obs}
\nonumber\\
&=&\frac{1}{\Delta \Omega_{\rm obs}} \int g^3 \delta_{\rm e} I_{\rm e}
e^{2 i \psi_{\rm obs}} d\Omega_{\rm obs} \, ,
\ee
where $\langle \cdot \rangle$ indicates the average over the image, the subindices ``obs'' and ``e'' refer, respectively, to quantities measured in the rest-frame of the observer and of the emitter, $\Delta \Omega_{\rm obs}$ is the total solid angle subtended by the disk on the sky, and the redshift factor $g = E_{\rm obs}/E_{\rm e}$ enters from the conservation of the quantity $I/E^3$ along the photon path, namely $I_{\rm obs}/E_{\rm obs}^3 = I_{\rm e}/E_{\rm e}^3$. We note that, in general, the component of the radiation that is initially completely polarized is detected on the observer's plane as partially polarized, because different points of the image have photons with different $\psi_{\rm obs}$. The total intensity at the detection point is
\be
\langle I_{\rm obs} \rangle &=&
 \frac{1}{\Delta \Omega_{\rm obs}} \int
g^3 I_{\rm e} d\Omega_{\rm obs} \nonumber\\
&=& \langle I^u_{\rm obs} \rangle + \langle I^p_{\rm obs} \rangle \, .
\ee
The observed averaged polarization degree is~\cite{lixin}
\be
\langle \delta_{\rm obs} \rangle &=& \, 
\frac{\sqrt{\langle Q_{\rm obs} \rangle^2 + \langle U_{\rm obs} 
\rangle^2}}{\langle I_{\rm obs} \rangle} \, , 
\ee
and the observed averaged polarization angle is determined from the following two relations~\cite{lixin}
\be
\sin \left( 2 \langle \psi_{\rm obs} \rangle \right) &=& \frac{\langle U_{\rm obs} 
\rangle}{\sqrt{\langle Q_{\rm obs} \rangle^2 + \langle U_{\rm obs} \rangle^2}} \, , \\
\cos \left( 2 \langle \psi_{\rm obs} \rangle \right) &=& \frac{\langle Q_{\rm obs} 
\rangle}{\sqrt{\langle Q_{\rm obs} \rangle^2 + \langle U_{\rm obs} \rangle^2}} \, .
\ee

With the above machinery, we can compute the spectrum of the polarization degree and of the polarization angle for a specific model. In our non-Kerr model, there are six basic parameters, namely the mass $M$, the distance $d$, the inclination angle of the disk with respect to the line of sight of the observer $i$, the mass accretion rate $\dot{M}$, the spin parameter $a_*$, the deformation parameter $\epsilon_3$. In what follows, we always assume $M = 10$~$M_\odot$ and an accretion luminosity to Eddington luminosity ratio of 0.1. The latter sets the mass accretion rate via $L = \eta \dot{M}$, where $\eta$ is the radiative efficiency in the Novikov-Thorne model (see the discussion in Section~3 in Ref.~\cite{lingyao}). We have checked the results of our code in the Kerr metric with those reported in Ref.~\cite{schnittman}. Fig.~\ref{fig1} shows the spectrum of the polarization degree (left panel) and the spectrum of the polarization angle (right panel) for a Schwarzschild BH (solid lines) and a Kerr BH with spin parameter $a_* = 0.9$ (dashed lines) when the viewing angle is $i = 45^\circ$, $60^\circ$, and $75^\circ$ (from bottom to top in the left panel, from top to bottom in the right panel). These plots can be compared with those in Fig.~2 in Ref.~\cite{schnittman} and the agreement is good. The apparent image of the accretion disk around a Kerr BH with spin parameter $a_* = 0.99$ is shown in Fig.~\ref{fig2}. The contour map shows the relative intensity of the total (namely polarized and unpolarized) radiation, $I_{\rm obs}/I_{\rm obs,max}$ (logarithmic scale). The black segments show the polarization of the radiation: the length of the segment is proportional to the polarization degree, while its orientation corresponds to that of the polarization vector. Our Fig.~\ref{fig2} can be compared with Fig.~1 in Ref.~\cite{schnittman} and the result is definitively similar.

\begin{figure*}
\begin{center}
\includegraphics[type=pdf,ext=.pdf,read=.pdf,width=8.7cm]{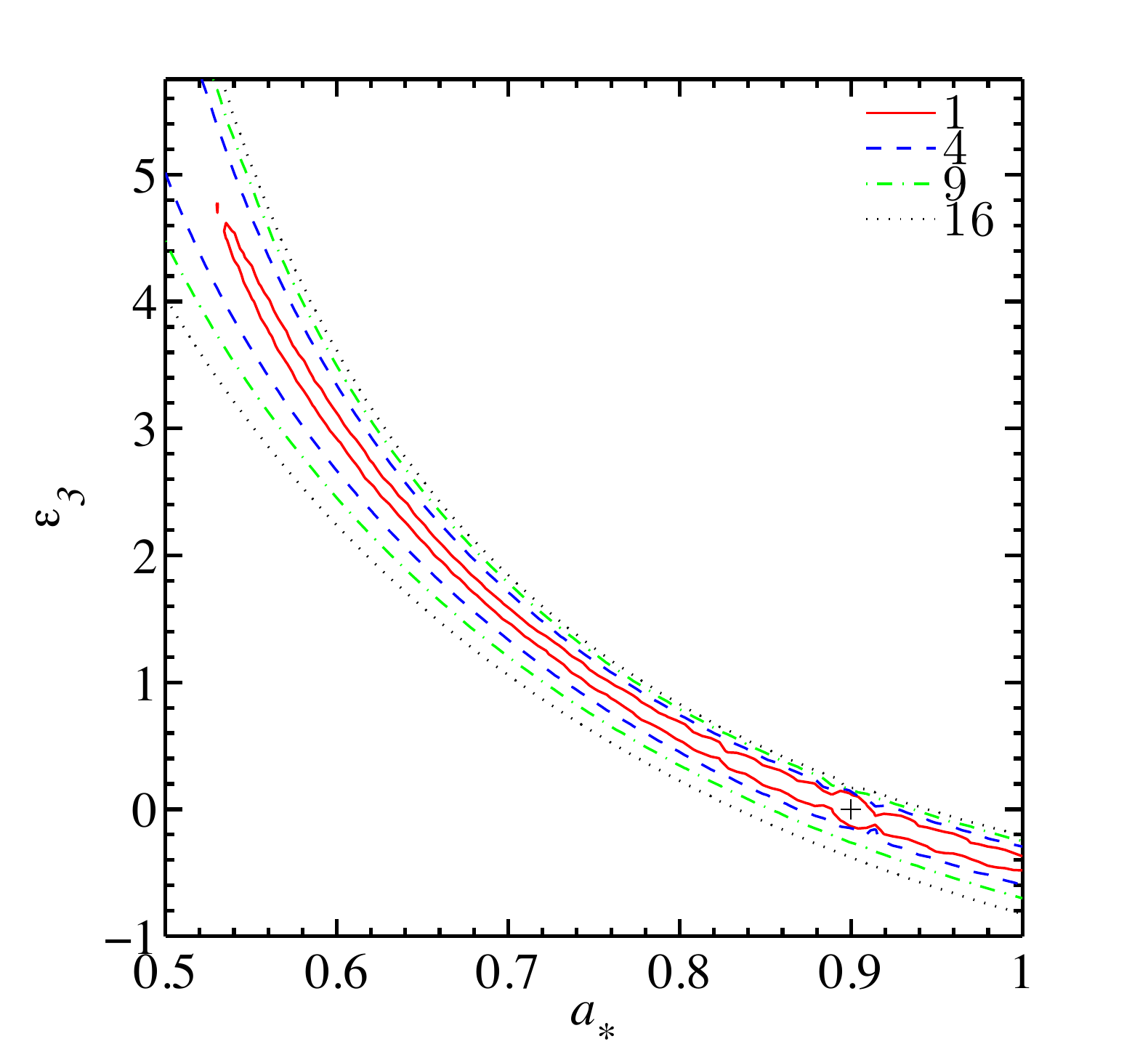}
\includegraphics[type=pdf,ext=.pdf,read=.pdf,width=8.7cm]{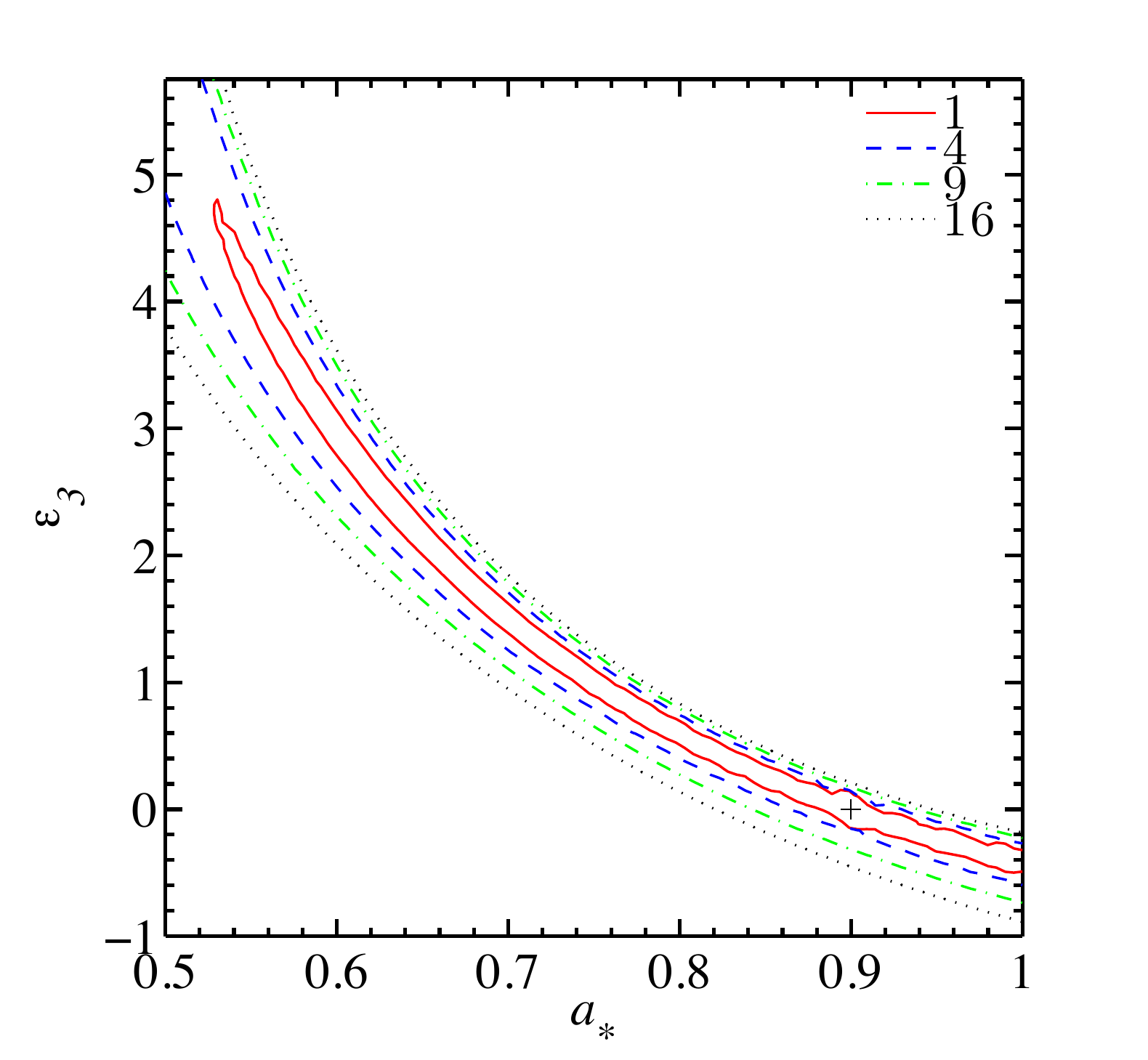}
\end{center}
\caption{$\Delta \chi^2$ contour lines on the plane spin parameter $a_*$ and deformation parameter $\epsilon_3$ from X-ray spectropolarimetric measurements. The reference model is a Kerr BH with spin parameter $a_* = 0.9$ and observed from a viewing angle $i=70^\circ$. In the left panel, all models use an inclination angle of $70^\circ$. In the right panel, the inclination angle is free in the fit. See the text for more details. \label{fig3}}
\vspace{0.5cm}
\begin{center}
\includegraphics[type=pdf,ext=.pdf,read=.pdf,width=8.7cm]{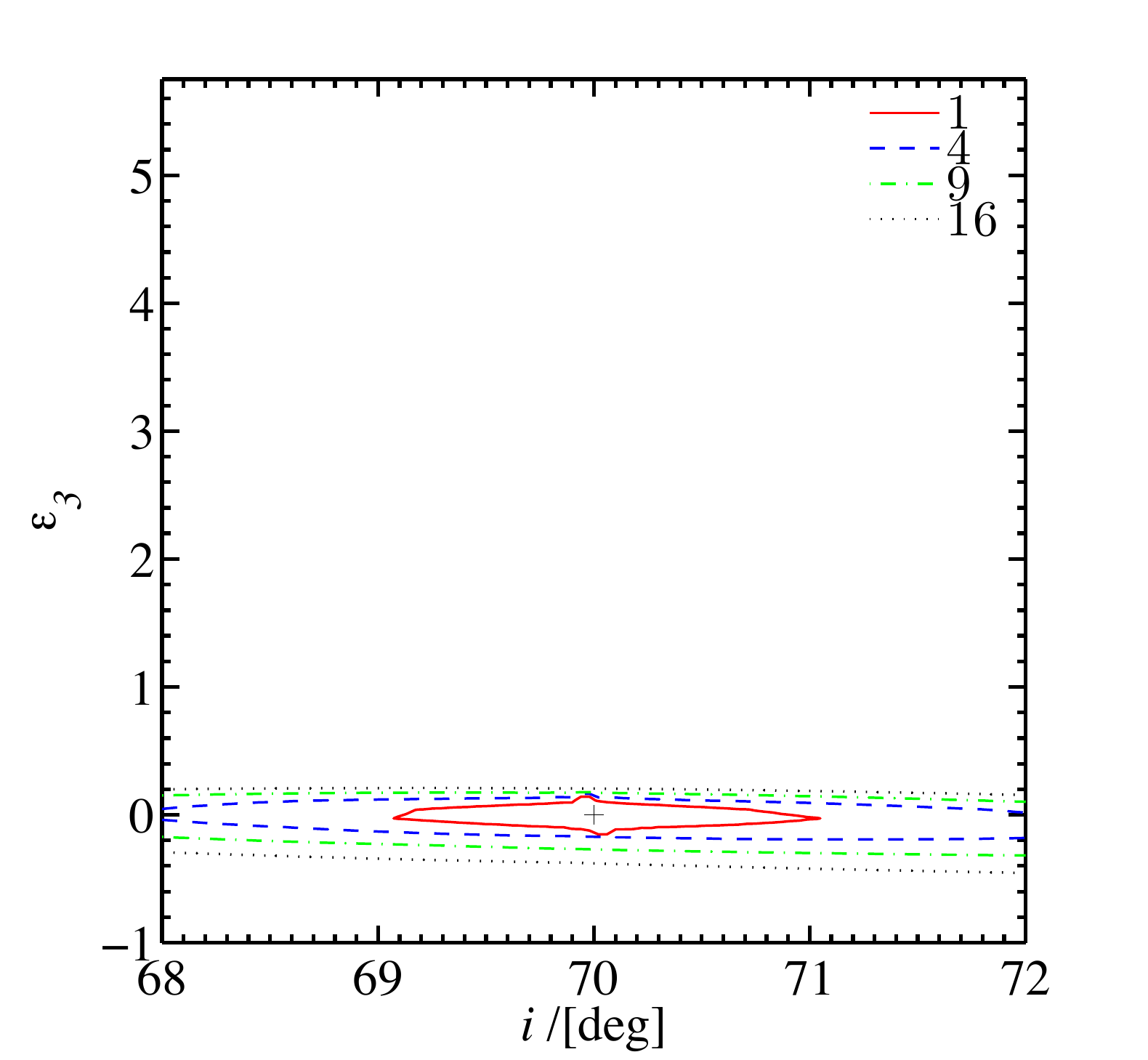}
\includegraphics[type=pdf,ext=.pdf,read=.pdf,width=8.7cm]{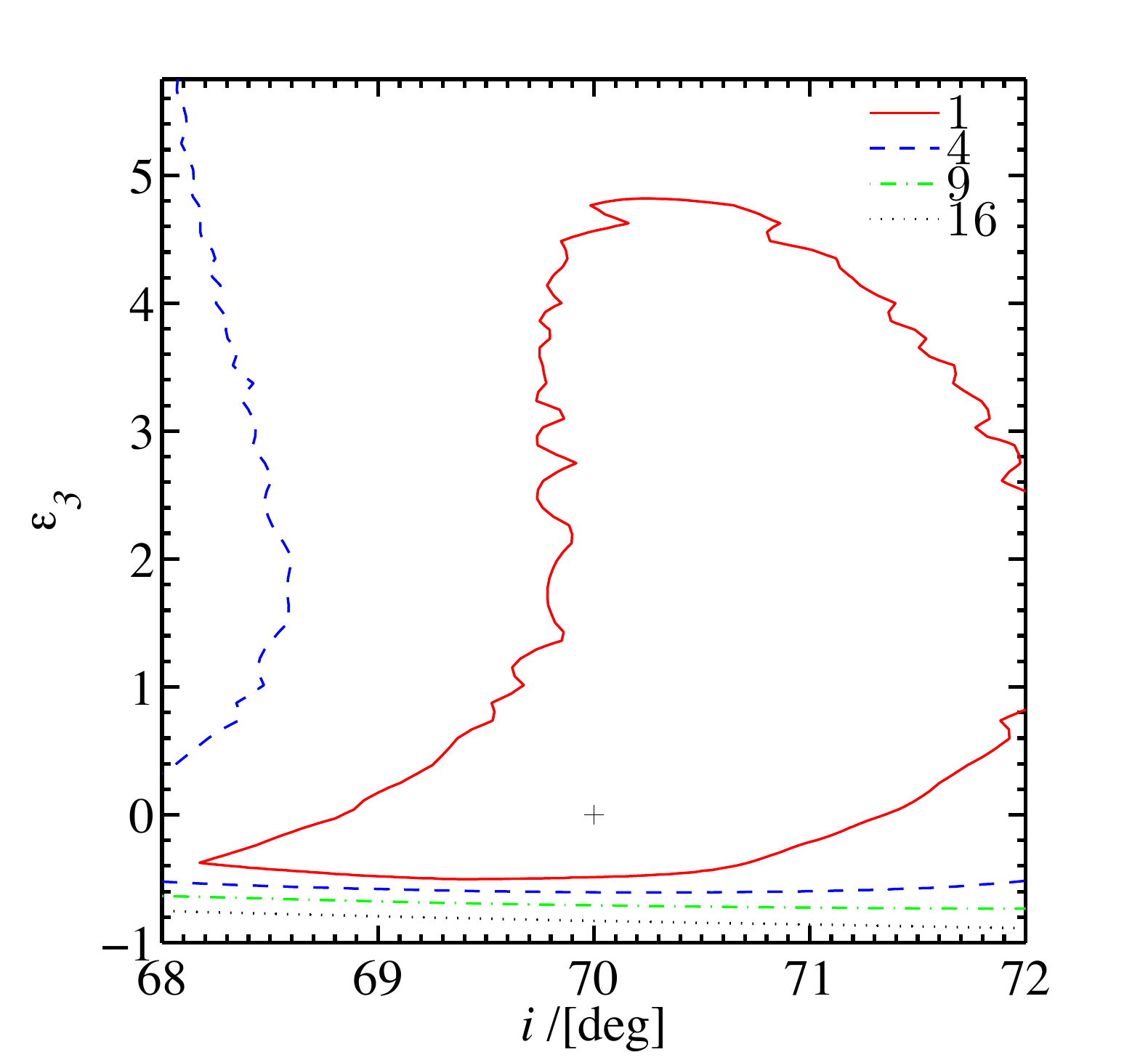}
\end{center}
\caption{As in Fig.~\ref{fig3}, on the plane inclination angle $i$ and deformation parameter $\epsilon_3$. In the left panel, all models use a spin parameter $a_* = 0.9$. In the right panel, the spin is free in the fit. See the text for more details. \label{fig4}}
\end{figure*}

\begin{figure*}
\begin{center}
\includegraphics[type=pdf,ext=.pdf,read=.pdf,width=8.7cm]{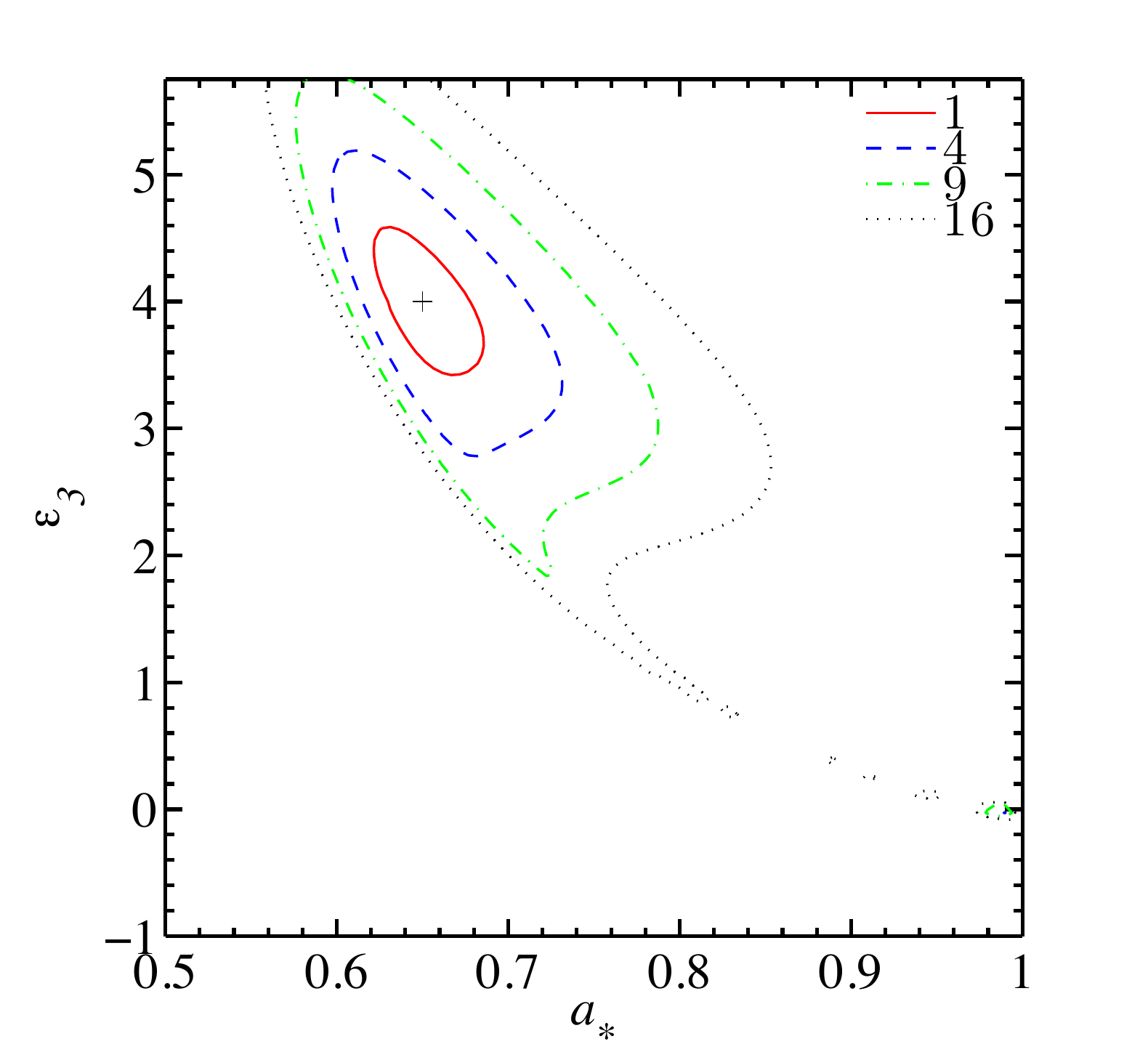}
\includegraphics[type=pdf,ext=.pdf,read=.pdf,width=8.7cm]{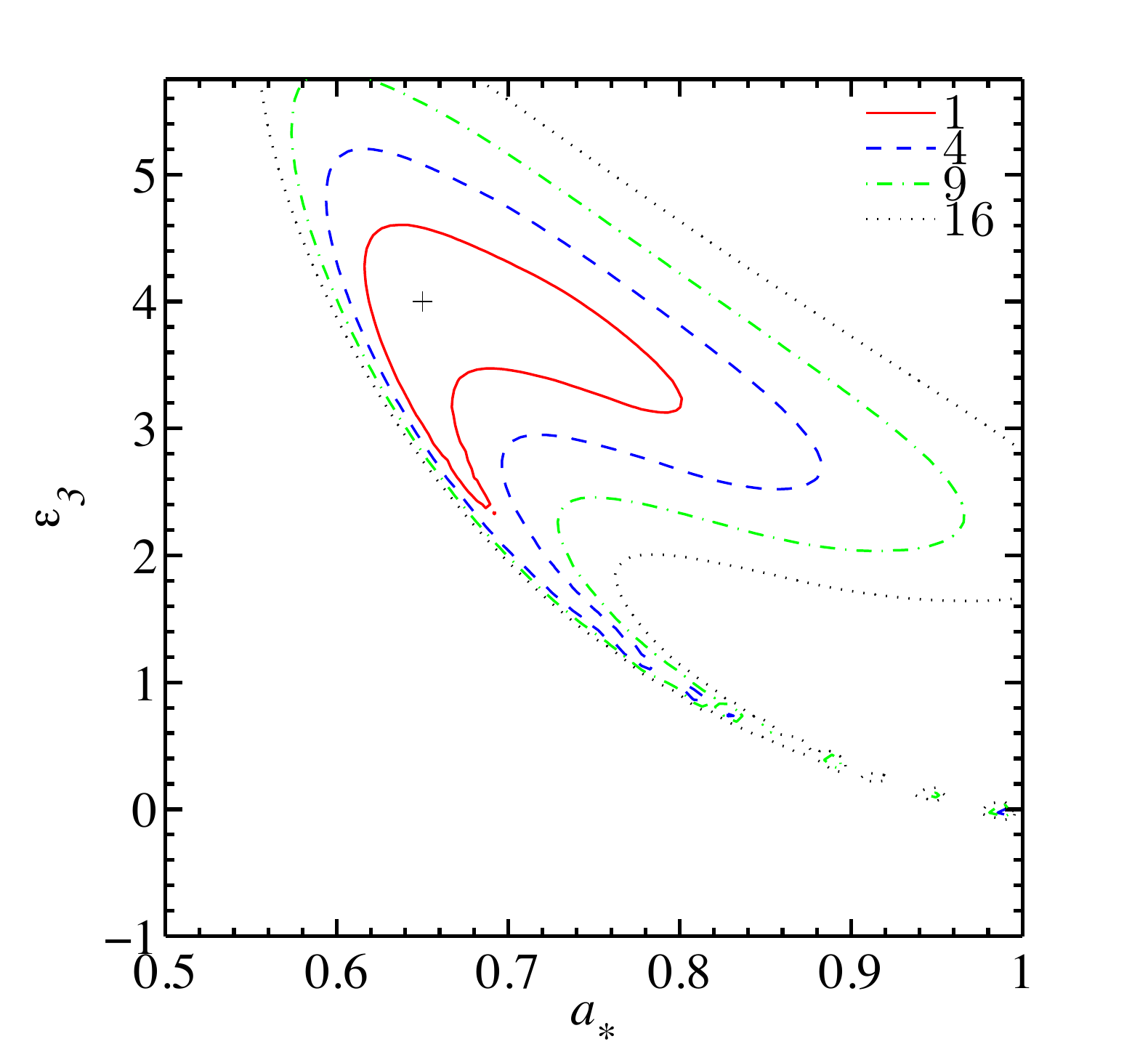}
\end{center}
\caption{$\Delta \chi^2$ contour lines on the plane spin parameter $a_*$ and deformation parameter $\epsilon_3$ from X-ray spectropolarimetric measurements. The reference model is a non-Kerr BH with spin parameter $a_* = 0.65$, deformation parameter $\epsilon_3 = 4$, and observed from a viewing angle $i=70^\circ$. In the left panel, all models use an inclination angle of $70^\circ$. In the right panel, the inclination angle is free in the fit. See the text for more details. \label{fig5}}
\vspace{0.5cm}
\begin{center}
\includegraphics[type=pdf,ext=.pdf,read=.pdf,width=8.7cm]{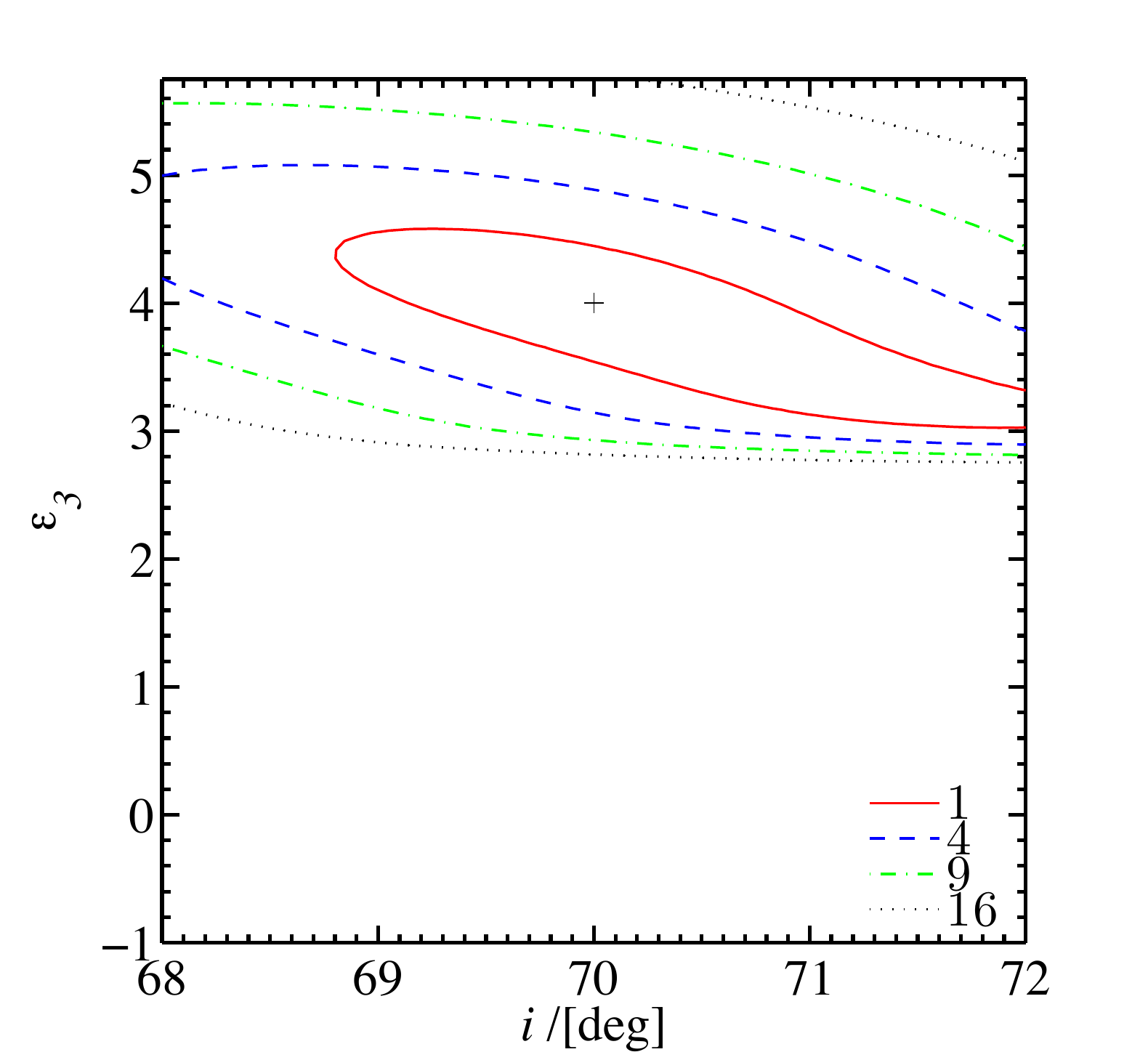}
\includegraphics[type=pdf,ext=.pdf,read=.pdf,width=8.7cm]{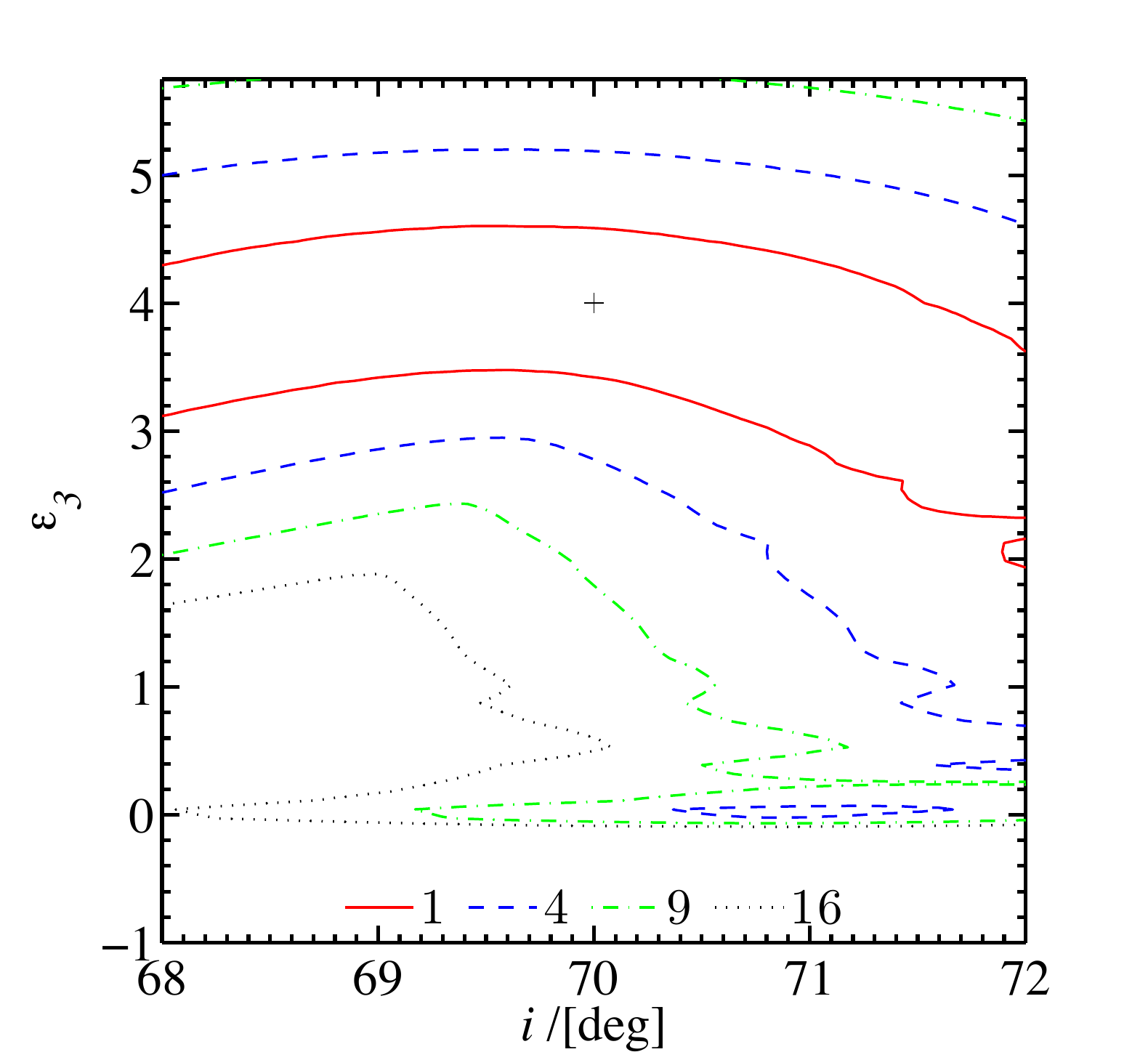}
\end{center}
\caption{As in Fig.~\ref{fig5}, on the plane inclination angle $i$ and deformation parameter $\epsilon_3$. In the left panel, all models use a spin parameter $a_* = 0.65$. In the right panel, the spin is free in the fit. See the text for more details. \label{fig6}}
\end{figure*}

\section{Tests of the Kerr metric} \label{s-4}

In this section, we want to figure out how X-ray spectropolarimetric measurements of the thermal spectrum of a thin disk (the so-called continuum component) can constrain the Kerr metric around a BH candidate, and, in particular, the parameter degeneracy. For a systematic study, we start considering a model with specific values of the spin parameter, deformation parameter, and inclination angle (reference model). This is compared with a set of Kerr and non-Kerr models with different values of $a_*$, $\epsilon_3$, and $i$ (template models) with the use of $\chi^2$. As done in Ref.~\cite{schnittman}, we define $\chi^2$ as 
\be
\hspace{-0.6cm}
\chi^2 (a_*, \epsilon_3, i) =
\sum_{k=1}^n \left[ \frac{\left( Q_k - Q_k^{\rm ref} \right)^2}{\Delta Q_k^2}
+ \frac{\left( U_k - U_k^{\rm ref} \right)^2}{\Delta U_k^2} \right] , 
\ee
where the summation is performed over $n$ sampling energies $E_k$, $Q_k$ and $U_k$ are the Stokes parameters of the template spectrum with parameters $a_*$, $\epsilon_3$, and $i$ in the energy bin $[E_k, E_k + \Delta E]$, while $Q_k^{\rm ref}$ and $U_k^{\rm ref}$ are the Stokes parameters of the reference spectrum in the energy bin $[E_k, E_k + \Delta E]$. The errors $\Delta Q_k$ and $\Delta U_k$ are assumed to be
\be
\Delta Q_k = \Delta U_k = \delta_{\rm min} 
I_k^{\rm ref} \sqrt{\frac{I^{\rm ref}_{\rm peak}}{I^{\rm ref}_k}} \, ,
\ee
where $\delta_{\rm min}$ is the minimum polarization sensitivity at the peak of the spectrum, $I_k^{\rm ref}$ is the total (polarized and unpolarized) intensity of the reference spectrum in the energy bin $k$, and $I^{\rm ref}_{\rm peak}$ is the total intensity of the reference spectrum at the peak. In this way we see how an observation can constrain the parameters of a putative source with the same values of the spin, the deformation, and the inclination angle as the reference model.

In this work, we assume that the energy range of the detector is 0.5-10~keV, the energy resolution is $\Delta E = 0.1$~keV, and that $\delta_{\rm min} = 0.003$. The results of our simulations are reported in Figs.~\ref{fig3}-\ref{fig6}. These plots show the contour levels $\Delta \chi^2 = \chi^2 - \chi^2_{\rm min} = 1$, 4, 9, and 16, where here $\Delta \chi^2 = \chi^2$ because $\chi^2_{\rm min} = 0$, since the minimum corresponds to the reference model without noise (the introduction of the noise would not change our results, of course). In the case of one degree of freedom, the 68.3\% C.L. designed as 1-standard deviation limit corresponds to $\Delta \chi^2 = 1$~\cite{lampton}. In the case of three degrees of freedom, we have $\Delta \chi^2 = 3.53$, 8.03, and 14.16, respectively for 68.3\%, 95.4\%, and $99.7\%$ (1-, 2-, 3-standard deviation limit).

As the first case, as reference model we consider a Kerr BH with spin parameter $a_* = 0.9$ and observed from an inclination angle $i = 70^\circ$. We note that a similar object would be a good source for our purpose, because the value of both the spin and the inclination angle are quite high and this should maximize relativistic effects, making the difference in the spacetime geometry more evident. Fig.~\ref{fig3} and Fig.~\ref{fig4} show, respectively, the constraints on the $(a_*,\epsilon_3)$ and $(i,\epsilon_3)$ planes. In the left panels, the third parameter is fixed, while in the right panels it is free in the fit and we minimize $\chi^2$ with respect to it. From Fig.~\ref{fig3}, we see that there is a strong correlation between the estimate of $a_*$ and $\epsilon_3$ and that it is very similar to that found in the case of the continuum-fitting method and the iron line analysis (see in particular Refs.~\cite{lingyao,cfmiron}). The possibility of an independent accurate measurement of $i$ is not very helpful to break the degeneracy between $a_*$ and $\epsilon_3$, and indeed the difference between the left and right plots is small. This conclusion is confirmed by Fig.~\ref{fig4}. An independent good measurement of $a_*$ would be enough to constrain $\epsilon_3$ (left panel), but without it large deviations from the Kerr geometry cannot be excluded (right panel).

Fig.~\ref{fig5} and Fig.~\ref{fig6} show the contour levels for a hypothetical non-Kerr BH with $a_* = 0.65$ and $\epsilon_3 = 4$. The inclination angle of the reference model is still $i = 70^\circ$. These plots confirm the conclusion reported in the previous paragraph. We should note that in Fig.~\ref{fig5} the allowed region is actually much longer than that apparently shown and that it extends to the Kerr models with $\epsilon_3 = 0$. The allowed region is very narrow for a fixed $\epsilon_3$, and therefore the numerical simulations miss it, but this can be understood by noting that both in the left and right panels there are some allowed ``islands'' near $a_* = 1$ and $\epsilon_3 = 0$. Once again, the Kerr metric could be tested with an independent estimate of $a_*$ (Fig.~\ref{fig6}, left panel), while it is not possible to do it if $a_*$ is free in the fit (Fig.~\ref{fig6}, right panel).

\section{Summary and conclusions} \label{s-c}

Current observations cannot unambiguously confirm whether astrophysical BH candidates are the Kerr BHs or general relativity because of a degeneracy among the parameters of the system: a non-Kerr compact object may mimic a Kerr BH with a different spin parameter and observed from a slightly different viewing angle. In the case of the shape of the thermal spectrum of thin disks, such a degeneracy is at the fundamental level and more accurate measurements cannot break it~\cite{cfm2}. Concerning the iron line profile, this technique is potentially more powerful: with the available X-ray data, the intrinsic Poisson noise of the source does not allow to break the degeneracy between the spin and possible deviations from the Kerr geometry, but future high-quality data can do it~\cite{jjc}, even if not for any kind of deformations~\cite{jjc2}.

The radiation of the thermal spectrum of thin accretion disks around BH candidates is inevitably polarized due to Thomson scattering of X-ray photons off free electrons in the dense atmosphere above the disk. The degree of polarization depends on the photon direction with respect to the normal of the disk surface, while the polarization vector is initially parallel to the plane of the disk. However, the strong gravitational field in the vicinity of BH candidates strongly affects the degree and the angle of polarization detected by an observer far from the object and polarization measurements can provide information about the metric. In this paper, we have studied the possibilities offered by X-ray spectropolarimetric measurements to test the Kerr geometry around BH candidates. Our targets are stellar-mass BH candidates in the high/soft state. This kind of measurements may be possible in the near future with the XTP~\cite{xtp} and the XIPE~\cite{xipe} missions.

As the first step in our program, we have considered a simple theoretical model, in which we neglect the effects of returning radiation and photon absorption by the disk's atmosphere, two ingredients that will be included in future developments. We have considered an hypothetical detector with an energy resolution of 0.1~keV in the 0.5-10~keV band and a detection sensitivity of the polarization degree at the peak of the spectrum of 0.003. As reference models, we have studied a Kerr BH with spin $a_* = 0.9$ and a non-Kerr BH with spin $a_* = 0.65$ and deformation parameter $\epsilon_3 = 4$. Both the reference models have an inclination angle $i = 70^\circ$. Our results can be summarized as follows:
\begin{enumerate}
\item We confirm the conclusions of Ref.~\cite{kraw}, namely that X-ray spectropolarimetric measurements cannot unambiguously test the Kerr metric because there is a strong correlation between the estimate of the spin and possible deviations from the Kerr solution. 
\item The correlation between these two parameters and the estimate of the inclination angle of the disk is instead modest, with the result that an independent accurate measurement of $i$ does not help to break the degeneracy between the spin and the deformation parameter.
\item The correlation between the estimates of the spin and the deformation parameter in X-ray spectropolarimetric measurements is very similar to those already found in the case of the continuum-fitting method and of the analysis of the iron K$\alpha$ line. This suggests that the possible combinations of the three measurements is not promising to break the degeneracy.
\item In the case of a plausible future polarization measurement, it seems that the capability to test the Kerr metric with this technique is roughly comparable to that offered by the continuum-fitting method and definitively worse than high quality data of the iron line.
\item If we relax the Kerr BH hypothesis, constraints on the disk's inclination angle with X-ray spectropolarimetric measurements become weaker.
\end{enumerate} 
The characteristic of our hypothetical detector are probably a little bit too optimistic for a first-generation of X-ray spectropolarimetric detectors, and therefore such a technique may not compete with the available ones in the near future. However, we must note that our results cannot be taken as conclusive. As shown in Ref.~\cite{schnittman}, the scattering of returning radiation can imprint its signature just above a few keV in the case of fast-rotating Kerr BHs. Since the effect is caused by the light bending in the vicinity of the object, it may provide further constraints and/or an independent information with respect to the thermal spectrum or the iron line. A final answer on the possibility of testing the Kerr metric with X-ray polarimetric data thus requires to properly include the effect of returning radiation, and this is left to future work.


\begin{acknowledgments}
We would like to thank Damiano Anselmi, Caigan Chen, and Matteo Guainazzi for useful comments and suggestions. This work was supported by the NSFC grant No.~11305038, the Shanghai Municipal Education Commission grant for Innovative Programs No.~14ZZ001, the Thousand Young Talents Program, and Fudan University.
\end{acknowledgments}


\begin{widetext}

\appendix

\section{Polarization calculations \label{s-a}}

We consider an observer at a distance $d$ from the BH and with an inclination angle $i$. His/her image plane is provided with a system of Cartesian coordinates $(X,Y,Z)$, as shown in Fig.~\ref{setup}. Another system of Cartesian coordinates $(x,y,z)$ is centered at the BH with the orientation reported in Fig.~\ref{setup}. The two Cartesian coordinates are related by 
\be
x = d \sin i - Y \cos i + Z \sin i \, , \quad
y = X \, , \quad
z = D \cos i + Y \sin i + Z \cos i \, .
\ee
Far from the BH, the Boyer-Lindquist spatial coordinates reduce to the usual spherical coordinates and they are related to $(x,y,z)$ by
\be
r = \sqrt{x^2 + y^2 + z^2} \, , \quad 
\theta = \arccos \left(\frac{z}{r}\right) \, , \quad 
\phi = \arctan \left(\frac{y}{x}\right) \, .
\ee

\subsection{Initial conditions}

The image of the accretion disk is provided by the photons hitting the image plane of the distant observer with 3-momentum perpendicular to the plane. In our numerical calculations, we start from photons on the image plane and we follow their trajectory and the evolution of their polarization vector till they hit the accretion disk.

Let us consider a photon at the position $(X_0, Y_0, 0)$, with 3-momentum ${\bf k}_0 = - k_0 \hat{Z}$ perpendicular to the image plane, and an auxiliary 3-vector ${\bf f}_0 = \hat{X}$. The initial conditions for the photon position are
\be
t_0 = 0 \, , \quad
r_0 = \sqrt{X_0^2 + Y_0^2 + d^2} \, , \quad
\theta_0 = \arccos \frac{Y_0 \sin i + d \cos i}{r_0} \, , \quad
\phi_0 = \arctan \frac{X_0}{d \sin i - Y_0 \cos i} \, ,
\ee
and for the photon 4-momentum are
\be
k^r_0 = - \frac{d}{r_0} |{\bf k}_0| \, , \quad
k^\theta_0 = \frac{\cos i - \left(Y_0 \sin i + d \cos i\right) 
\frac{d}{r_0^2}}{\sqrt{X_0^2 + (d \sin i - Y_0 \cos i)^2}} |{\bf k}_0| \, , \quad
k^\phi_0 = \frac{X_0 \sin i}{X_0^2 + (d \sin i - Y_0 \cos i)^2} |{\bf k}_0| \, ,  
\ee
and $k^t_0 = \sqrt{\left(k^r_0\right)^2 + r^2_0  \left(k^\theta_0\right)^2 + r_0^2 \sin^2\theta_0  (k^\phi_0)^2}$ follows from the condition $g_{\mu\nu}k^\mu k^\nu = 0$ with the metric tensor of a flat space-time. The initial conditions for the auxiliary vector are 
\be
f^t_0 = 0 \, , \quad
f^r_0 = \frac{X_0}{r_0} \, , \quad
f^\theta_0 = \frac{d \cos i + Y_0 \sin i}{\sqrt{X^2_0 + \left(d \sin i - Y_0 \cos i\right)^2}} 
\frac{X_0}{r_0^2}\, , \quad
f^\phi_0 = \frac{d \sin i - Y_0 \cos i}{X^2_0 + \left(d \sin i - Y_0 \cos i\right)^2} \, .
\ee

\subsection{Propagation in non-Kerr spacetimes}

The photon trajectory is found by solving the geodesic equations
\be
\frac{d^2 x^\mu}{d\lambda^2} + \Gamma^\mu_{\nu\sigma} 
\frac{dx^\nu}{d\lambda} \frac{dx^\sigma}{d\lambda}= 0
\ee
and the auxiliary vector is parallel transported along the photon geodesic
\be
\frac{df^\mu}{d\lambda} = - \Gamma^\mu_{\nu\sigma} f^\nu \frac{dx^\sigma}{d\lambda} \, .
\ee
In order to have numerical errors under control, at any step we check that the following relations are satisfied: $k^\mu k_\mu = 0$, $f^\mu f_\mu = 1$, $f^\mu k_\mu = 0$.

\begin{figure}
\begin{center}
\includegraphics[type=pdf,ext=.pdf,read=.pdf,width=9cm]{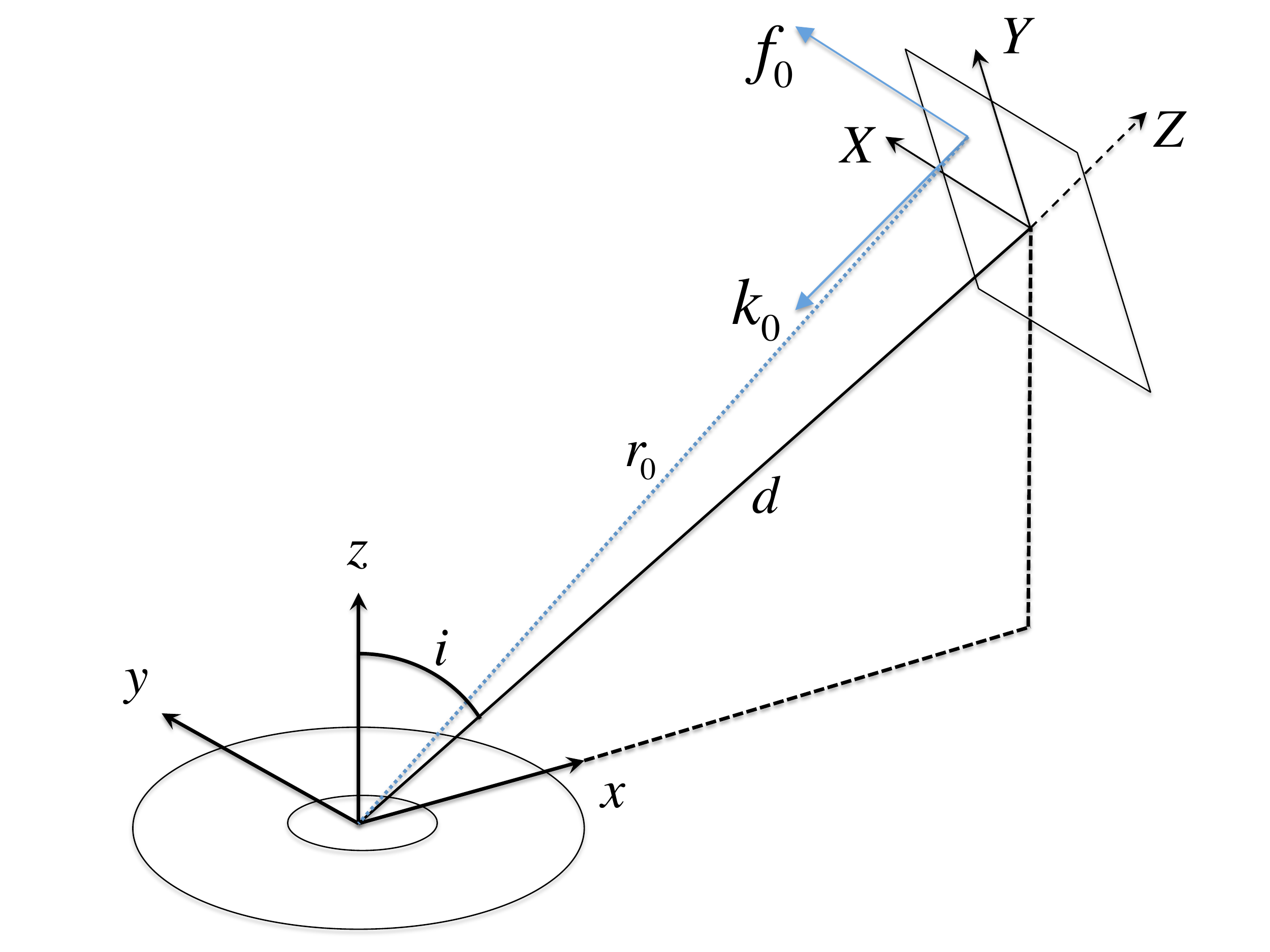}
\end{center}
\caption{Set-up of the system. The Cartesian coordinates $(x,y,z)$ are centered at the BH. The image plane of the distant observer is located at the distant $d$ from the BH, with an inclination angle $i$, and it is provided with a system of Cartesian coordinates $(X,Y,Z)$. In the calculations, a photon starts from the image plane with initial position $(X_0, Y_0, 0)$, initial 3-momentum ${\bf k}_0 = - k_0 \hat{Z}$ perpendicular to the image plane, and auxiliary 3-vector ${\bf f}_0 = \hat{X}$ perpendicular to its 3-momentum and parallel to the $X$-axis. See the text for more details.}
\label{setup}
\end{figure}

\subsection{Coordinate transformations at the emission point}

When the photon hits the disk, we have to determine: $i)$ the angle $\vartheta$ between the normal to the disk and the direction of propagation of the photon as measured in the rest frame of the accreting gas, $ii)$ the orientation of the auxiliary vector as measured in the rest frame of the accreting gas. The angle $\vartheta$ is necessary to obtain the polarization degree. The orientation of the auxiliary vector is compared to that of the polarization vector expected from the emerging radiation. Since the angle between the two vectors is conserved along the photon path, it is the same at the emission and at the detection points, and therefore we can immediately determine the angle of the polarization vector measured far from the BH.

Our spacetimes are stationary, axisymmetric, and asymptotically flat. We write the line element in the following form
\be
ds^2 = - e^{2\nu} dt^2 + e^{2\mu} dr^2 + e^{2\lambda} d\theta^2 + e^{2\sigma} \left(d\phi - \omega dt\right)^2 \, .
\ee
The covariant basis vectors in the locally non-rotating frame is
\be
E_\mu^{(t)} = \Big( \, e^\nu \, , \, 0 \, , \, 0 \, , \, 0 \, \Big) \, , \;
E_\mu^{(r)} = \Big( \, 0 \, , \, e^\mu \, , \, 0 \, , \, 0 \, \Big) \, , \;
E_\mu^{(\theta)} = \Big( \, 0 \, , \, 0 \, , \, e^\lambda \, , \, 0 \, \Big) \, , \;
E_\mu^{(\phi)} = \Big( \, - e^\sigma \omega \, , \, 0 \, , \, 0 \, , \, e^\sigma \, \Big) \, .
\ee
The gas 4-velocity in the locally non-rotating frame is $u^{(\mu)} = E_\nu^{(\mu)} u^\nu$ and therefore the velocity of the gas with respect to the locally non-rotating frame is
\be
v = \frac{u^{(\phi)}}{u^{(t)}} = \left(\Omega - \omega\right) e^{\sigma - \nu} \, ,
\ee
where $\Omega = u^\phi/u^t$ is the angular velocity of the gas as measured by the distant observer. The covariant basis vectors in the rest-frame of the gas are related to those in the locally non-rotating frame by the following Lorentz transformation
\be
\hat{E}_\mu^{(t)} = \gamma \left( E_\mu^{(t)} - v E_\mu^{(\phi)} \right) \, , \quad
\hat{E}_\mu^{(r)} = E_\mu^{(r)} \, , \quad
\hat{E}_\mu^{(\theta)} = E_\mu^{(\theta)} \, , \quad
\hat{E}_\mu^{(\phi)} = \gamma \left( - v E_\mu^{(t)} + E_\mu^{(\phi)} \right) \, .
\ee
where $\gamma = \left(1 - v^2\right)^{-1/2}$ is the Lorentz factor. Eventually, we can write the photon 4-momentum and the polarization 4-vector in the rest-frame of the gas
\be
\hat{k}^{(a)} = \hat{E}_\mu^{(a)} k^\mu \, , \quad 
\hat{f}^{(a)} = \hat{E}_\mu^{(a)} f^\mu \, .
\ee

\subsection{Calculation of $\vartheta$ and $\psi$}

The polarization degree $\delta$ depends on the angle $\vartheta$ between the normal to the disk surface and the direction of propagation of the photon. Once we know $\vartheta$, we can find the corresponding $\delta$ from the table in Ref.~\cite{chandra}. Since the emission is on the equatorial plane, we have
\be
\cos\vartheta = \frac{\left| \hat{k}^{(\theta)} \right| }{\sqrt{\left(\hat{k}^{(r)}\right)^2 + \left(\hat{k}^{(\theta)}\right)^2 + \left(\hat{k}^{(\phi)}\right)^2}} =  \frac{\left| \hat{k}^{(\theta)} \right|}{\hat{k}^{(t)}} \quad \Rightarrow
\quad \vartheta = \arccos \left(\frac{\left| \hat{k}^{(\theta)} \right|}{\hat{k}^{(t)}}\right) \, .
\ee

For the calculation of the polarization angle, we proceed in the following way. The polarization vector of the radiation emerging from the disk, say $\hat{h}^{(a)}$, is oriented perpendicular to the propagation direction of the photons and parallel to the disk (vanishing $\theta$-component). We choose the gauge in which the $t$-component of the polarization vector vanishes and, from the normalization condition $\hat{h}^{(a)} \hat{h}_{(a)} = 1$ and the orthogonality with the photon momentum $\hat{h}^{(a)} \hat{k}_{(a)} = 0$, we find
\be
\hat{h}^{(r)} = \frac{\hat{k}^{(\phi)}}{\sqrt{\left(\hat{k}^{(r)}\right)^2 + \left(\hat{k}^{(\phi)}\right)^2}} \, , \quad \hat{h}^{(\phi)} = - \frac{\hat{k}^{(r)}}{\sqrt{\left(\hat{k}^{(r)}\right)^2 + \left(\hat{k}^{(\phi)}\right)^2}}
\ee
Before comparing $\hat{h}^{(a)}$ with $\hat{f}^{(a)}$, we need to use the same gauge for the two vectors and we thus remove the $t$-component in $\hat{f}^{(a)}$, namely
\be
\hat{f}^{(a)} \rar \hat{f}'^{(a)} = \hat{f}^{(a)} - \alpha \hat{k}^{(a)} \, ,
\ee
\end{widetext}
where $\alpha = \hat{f}^{(t)}/\hat{k}^{(t)}$. We define the angle $\xi$ as
\be
\xi = 
\left\{
\begin{array}{ll}
\arccos{ \left(f^{\hat{r}} h^{\hat{r}} + f^{\hat{\phi}} h^{\hat{\phi}} \right)} & \textrm{if $f^{\hat{\theta}} > 0$}	\\
\arccos{ \left(- f^{\hat{r}} h^{\hat{r}} - f^{\hat{\phi}} h^{\hat{\phi}} \right)} & \textrm{if $f^{\hat{\theta}} \le 0$}
\end{array}
\right.
\ee
and the polarization angle detected by the distant observer is
\be
\psi = 
\left\{
\begin{array}{ll}
\frac{\pi}{2} - \xi & \textrm{if $\xi < \frac{\pi}{2}$}	\\
\frac{3 \pi}{2} - \xi & \textrm{if $\xi \ge \frac{\pi}{2}$}
\end{array}
\right. \, .
\ee


\end{document}